\documentclass[10pt,twocolumn]{article}
\usepackage[utf8]{inputenc}
\usepackage{amssymb}
\usepackage{amsmath}
\usepackage{graphicx}
\usepackage[english]{babel}
\newtheorem{myprop}{Proposition}
\newtheorem{mylemma}{Lemma}
\newtheorem{myso}{Corollary}
\newtheorem{mytheorem}{Theorem}
\title{Zipf and non-Zipf laws for homogeneous Markov chain}
\author{V.V.~Bochkarev and E.Yu.~Lerner}
\date{}
\begin{document}
\maketitle

\begin{abstract}
Let us consider a homogeneous Markov chain with discrete time and with a finite set of states $E_0,\ldots,E_n$ such that the state $E_0$ is absorbing, states $E_1,\ldots,E_n$ are nonrecurrent. The goal of this work is to study frequencies of trajectories in this chain, i.e., ``words'' composed of symbols $E_1,\ldots,E_n$ ending with the ``space''~$E_0$. 

Let us order words according to their probabilities; denote by $p(t)$ the probability of the $t$th word in this list. In this paper we prove that in a typical case the asymptotics of the function $p(t)$ has a power character, and define its exponent from the matrix of transition probabilities. If this matrix is block-diagonal, then with some specific values of transition probabilities the power asymptotics gets (logarithmic) addends. But if this matrix is rather sparse, then probabilities quickly decrease; namely, the rate of asymptotics is greater than that of the power one, but not greater than that of the exponential one. We also establish necessary and sufficient conditions for the exponential order of decrease and obtain a formula for determining the exponent from the transition probability matrix and the initial distribution vector.
\end{abstract}
\medskip 


\section{Introduction.}

In recent time, the nature of power laws and spheres of their use became of interest in applications~\cite{durett, mitzenmacher, newman}. For real networks one has proposed several models describing the occurrence of the power law; the most known one is the preferential attachment model~\cite{albert}. In linguistics, mechanisms of the occurrence of Zipf and Heaps laws were thoroughly studied in the time of B.~Mandelbrot~\cite{mand1, mand2}. Papers containing empirical studies and mathematical models also appear regularly nowadays (see, for example,~\cite{Lu} and references therein; for the mathematical motivation of this paper see~\cite{Mit}). However, there are no commonly accepted explanations of the fact that in reality with some values of parameters the power law does not adequately describe processes under consideration~\cite{newman}. Here we try to answer this question, considering probabilities of the occurrence of various trajectories in a homogeneous Markov chain.

Our model has occurred when studying a huge data set of the Google Books repository~\cite{google}. Usually one describes frequencies of words occurrences with the help of a power law asymptotics~\cite{Baayen}. But note that the power law is irrelevant in hieroglyphic scripts \cite{Lu}. 

As the initial model explaining the power law of decrease of frequencies of occurrences of English words we consider the model of the word generation process consisting in the sequential independent random addition of various symbols (letters and the space), each of which has a fixed probability (the monkey model). This model has a long history, but the power character of the asymptotics of the sorted list of word frequencies has been strictly justified for it only recently~\cite{Mit, VovaAndI}.

In this paper we study one natural generalization of this model, namely, the model with the Markov connection of neighboring symbols. Such model was studied by B.~Mandelbrot~\cite{mand2}; however, he has mainly considered a particular case of the occurrence of the power asymptotics. As appeared, in dependence of the matrix of transition probabilities, the ordered list of frequencies of all possible trajectories of a Markov chain can have essentially different asymptotics.

Thus, let us consider a homogeneous Markov chain with discrete time and with a finite set of states $E_0,\ldots,E_n$ such that
\begin{equation}
\label{MC}
\begin{array}{c}
\text{the state $E_0$ is absorbing,}\\
\text{states $E_1,\ldots,E_n$ are nonrecurrent}
\end{array}
\end{equation}
(see \cite{feller, kelbert} for the terminology and equivalent statements given below). The goal of this work is to study frequencies of trajectories in this chain, i.e., ``words'' composed of symbols $E_1,\ldots,E_n$ ending with the ``space''~$E_0$. 

Let us order words (trajectories) according to their probabilities; denote by $p(t)$ the probability of the $t$th word in this list. In this paper we prove that in a typical case the asymptotics of the function $p(t)$ has a power character, and define its exponent from the matrix of transition probabilities. If this matrix is block-diagonal, then with some specific values of transition probabilities the power asymptotics gets (logarithmic) addends. But if this matrix is rather sparse, then probabilities quickly decrease; namely, the rate of asymptotics is greater than that of the power one, but not greater than that of the exponential one. We also establish necessary and sufficient conditions for the exponential order of decrease and obtain a formula for determining the exponent from the transition probability matrix and the initial distribution vector.

\section{The exact statement of main result}

Let $P_0$ be a (stochastic) transition probability matrix of the Markov chain mentioned in the last but one paragraph, and let $P$ be its (substochastic) submatrix corresponding to states $E_1,\ldots,E_n$. Denote by $G_0$ the directed pseudograph with the set of vertices $\{0,\ldots,n\}$, whose arcs $(i,j)$ are defined by inequalities $p_{ij}>0$. Conditions~(\ref{MC}) are equivalent to the fact that the graph~$G_0$ is (weakly) connected, and $\{ 0\}$ is the only collection of vertices that has no arcs leading to its complement. Let $G$ be the subgraph of the graph~$G_0$ with the set of vertices $\{1,\ldots,n\}$ including all arcs of the initial graph~$G_0$ between these vertices ({\it the subgraph generated by vertices} $\{1,\ldots,n\}$). Let~$H$ be a subgraph of the graph~$G_0$ generated by some set of vertices. Then we denote by $P_H$ the corresponding submatrix of the matrix~$P_0$:$P_H=(p_{ij})_{i,j\in V(H)}$. Thus, for example, $P_G \equiv P$. In addition, we set $P_H(\beta)= (p_{ij}^\beta)_{i,j\in V(H)}$.

Recall that a {\it strongly connected component} is a maximal complete subgraph such that any pair of its vertices is mutually connected. Denote by $G'$ the digraph obtained from the graph~$G_0$ by identifying vertices and arcs that belong to all strongly connected components of the initial graph~$G_0$ (in~\cite{harary} this graph is called the {\it condensation}). In this paper, the graph $G'$ is connected and $0$ is the only vertex having no outgoing arcs. Recall that \cite{harary} the graph $G'$ is acyclic.

We denote by $a=(a_0,\dots,a_n)$ the initial distribution of probabilities on the state set. Without loss of generality, we assume that 
\begin{equation}
\label{cond}
\begin{array}{c}
\text{there are no states with zero probability}\\
\text{of reaching them at any time moment.}
\end{array}
\end{equation}

In what follows we sometimes deal with initial distributions, for which condition~(\ref{cond}) is not assumed to be fulfilled; we specify all such cases separately.

Let us associate an arbitrary {\it path}~$c=(i_1,\ldots,i_m)$ in the graph~$G_0$ with the weight $\widetilde{\mathop{\rm Pr}}(c)= p_{i_1 i_2} \ldots p_{i_{m-1} i_{m}}$. Instead of a path in the graph, it is often more convenient to consider an ordered set of states of the chain $w=(E_{i_1},\ldots,E_{i_m})$. We call this set a {\it word}, if $a_{i_1}>0$, $E_{i_m}=E_0$, and $E_{i_{m-1}}\ne E_0$. In other words, we understand a word as a sequence of states reached by the system from the start of the walk till the absorption by the state~$E_0$. We determine the word probability~$\mathop{\rm Pr}(w)$, taking into account the initial distribution:
\begin{equation}
\label{PrWord}
\mathop{\rm Pr}(w)=a_{i_1}p_{i_1 i_2} \ldots p_{i_{m-1} i_m}.
\end{equation}
One can easily prove that the set of all words with the measure $\mathop{\rm Pr}$ forms a discrete probability space (i.e., the sum of probabilities of all words equals one).

We understand the {\it length~$L$ of a word}~$w$ as the number of states in it, excluding the last absorbing state~$E_0$. We also denote by $C$ the \textit{set of all simple cycles} in the graph~$G$.

Let us sort all words in the nonincreasing order of their probabilities. Evidently, both the value $p(t)=\mathop{\rm Pr}(w_t)$ (the probability of the $t$th word in this ordered list) and the ``inverse'' to it function $Q(q)$, $q\in(0,1]$, (that equals the number of words whose probability is less than~$q$) are defined. We are interested in the asymptotics of the function~$p(t)$ for $t\to\infty$ (or, equivalently, that of the function $Q(q)$ for $q\to 0$).

We use the standard $O$-symbolics, namely, we denote by $\Theta$ the asymptotic order and we do by $\Omega$ the lower estimate of the order (\cite[section~9.2]{knuth}).

\medskip

\begin{figure}
\begin{center}
\includegraphics[width=0.5\textwidth]{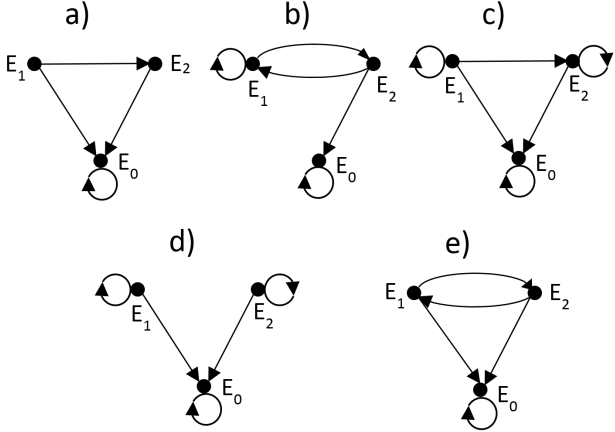}\quad
\caption{Examples of graphs~$G_0$ of a Markov chain with three states $E_0, E_1, E_2$ (the vertex that corresponds to the absorbing state $E_0$ is pictured at the bottom). In case~a) the function $p(t)$ is finitary. In case~b) the function $p(t)$ has a power asymptotics. In case~c) the function $p(t)$ decreases slower than any exponential function, but faster than any power one. In cases~d) and~e) the function $p(t)$ has an exponential asymptotics. Note that the classification depends only on the graph~$G$ (the upper part of the figure), provided that states~$E_1$ and $E_2$ are nonrecurrent.}
\end{center}
\label{lernerPic1}
\end{figure}

\begin{figure}
\begin{center}
\includegraphics[width=0.33\textwidth]{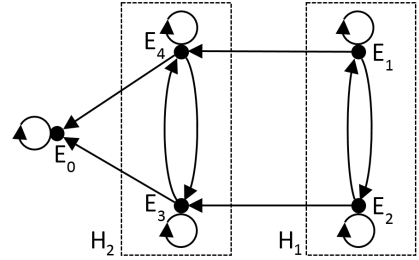}\quad 
\caption{An example of the graph~$G_0$ of a Markov chain with five states $E_0,E_1,E_2,E_3,E_4$. The function~$p(t)$ is bounded by two functions having a power asymptotics, however, their degrees are different (arbitrarily close). The function~$p(t)$ itself does not necessarily have a power asymptotics, if matrices of transition probabilities of graphs $H_1$ and $H_2$ coincide or so do the corresponding exponents $\beta$.}
\end{center}
\label{lernerPic2}
\end{figure}

\begin{mytheorem}
\label{main}
Three cases are possible:
\begin{itemize}
\item[{\rm\bf 1.}] If the graph~$G$ is acyclic, then the function $p(t)$ is finitary (i.e., the number of all possible words is finite).
\item[{\rm\bf 2.}] If the graph~$G$ contains a vertex which is common for two different simple cycles, then $p(t)=\Omega(t^{-1/\beta})$, where $\beta$ is a real number, with which the maximal modulo eigenvalue of the matrix $P_G(\beta)$ equals one. Note that such $\beta$ exists, is unique, and belongs to the interval $(0,1)$. Moreover, $p(t)=o(t^{-1/\beta'})$ for any $\beta'>\beta$. In addition, the exact power order (i.e., the equality $p(t)=\Theta(t^{-1/\beta})$) is attained if and only if any simple path in the graph~$G'$ contains at most one vertex (a strongly connected component~$H$ of the graph~$G$) such that the matrix $P_H(\beta)$ has the unit eigenvalue.
\item[{\rm\bf 3.}] If the graph~$G$ contains cycles, and each vertex of the graph~$G$ belongs to no more than one simple cycle, then $p(t)=\Omega(\alpha^t)$ and $p(t)=o(t^{-\lambda})$, where $\lambda$ is any positive value, while $\alpha\in (0,1)$ is some constant depending on the matrix $P$ (i.e., $p(t)$ decreases faster than any power function, but slower than a certain exponential one). This item includes a specific case, when the function~$p(t)$ decreases exponentially, namely, $p(t)=O\left(\exp{(-\kappa t)}\right)$ for some $\kappa>0$ if and only if any path in the graph~$G$ contains 
vertices of no more than one cycle. In this case we have $p(t)=\Theta(\exp{(-\nu t)})$; here $\nu$ is determined by the formula $1/\nu=-\sum_{c\in C} k(c)/\ln \widetilde{\mathop{\rm Pr}}(c)$, where $k(c)$ is the number of various words with nonrepeating states (simple paths that begin at vertices $v$ such that $a_v>0$) in the graph $G_0$ going through certain vertices of the cycle~$c$.
\end{itemize}
\end{mytheorem}

{\bf Remark~1.} The first item of the Theorem is trivial (we give it here only for the sake of completeness). It follows from the fact that in an acyclic graph the length of any word does not exceed~$n$.
\medskip

{\bf Remark~2.} The order of the exponential asymptotics (as distinct from the power case) depends not only on the matrix of transition probabilities, but also on the set of states $v$ such that $a_v>0$.

\medskip 

{\bf Examples.} The graph shown in diagram~b) in Fig.~1 has only one strongly connected component with vertices $\{1,2\}$ (we do not take into account the trivial cycle from the absorbing state to itself). This component contains cycles~$(1,2,1)$ and $(1,1)$, therefore, the function~$p(t)$ has a power asymptotics. For example, if all probabilities of transitions from states~$E_1$ and $E_2$ equal $1/2$, then one can easily calculate that $\beta=\log_2 (1+\sqrt{5})/2$. The graph shown in Fig.~\ref{lernerPic2} has two strongly connected components~$H_1$ and~$H_2$ (we do not take into account the trivial cycle 
from the absorbing state to itself), and both of them belong to one and the same path in the graph~$G'$. Thus, all conditions of Theorem~\ref{main}.2 are fulfilled. If probabilities of all transitions from states~$E_1,E_2,E_3,E_4$ equal $1/3$, then one can easily calculate that $\beta=\log_3 2$. With this value of $\beta$ matrices $P_{H_1}(\beta)$ and $P_{H_2}(\beta)$ have the unit eigenvalue (all their elements equal 1/2). Therefore the power asymptotics does not take place, i.e., 
$p(t)=\Omega( t^{-\log_2 3})$ and $p(t)=o(t^{-\delta})$ for any $\delta<\log_2 3$, but $p(t)\neq\Theta(t^{-\log_2 3})$.

The graph shown in diagram~c) in Fig.~1 contains two simple cycles-loops, and in the graph~$G$ there is a path going through all vertices, therefore, $p(t)$ decreases slower than an exponential function, but faster than a power one. The graph shown in diagram~d) in Fig.~1 contains two analogous cycles, but in the graph~$G$ there is no path described in the previous example; this means that the decrease of the function~$p(t)$ has an exponential asymptotics. Note that $k(c)=1$ for each of cycles. The graph shown in diagram~e) has one simple cycle, and the asymptotics is also exponential. Assume that $a_1>0$ and $a_2>0$; then $k(c)=4$, the four desired words with nonrepeating states are $(E_1,E_0), (E_2,E_0), (E_1,E_2,E_0), (E_2,E_1,E_0)$. Now assume that for Markov chains with graphs shown in diagrams~d)~and~e) all probabilities of transitions from states~$E_1,E_2$ equal $1/2$; then one can easily calculate that in both cases $\nu=\ln\sqrt{2}$.

\medskip 

{\bf Remark~3.} As was proved earlier~\cite{Mit,VovaAndI}, if states are chosen independently and the probability of each one is $p_i$, $i=0,\ldots,n$, then for $n>1$ the function $p(t)$ has a power asymptotics; its exponent determined from the equation~$\sum_{i=1}^n p_i^\beta=1$ equals~$1/\beta$. This is a particular case of Theorem~\ref{main}.2, where the matrix~$P$ consists of nonzero elements and has equal rows. Raising all elements of the matrix~$P$ to the power~$\beta$, we obtain a stochastic matrix; it is well known that the maximal eigenvalue of a stochastic matrix equals one.

\section{Spectral properties of substochastic matrices.}

Prior to proving Theorem~\ref{main}.2, let us prove the unique existence of the exponent~$\beta$ in this case. Consider an arbitrary substochastic matrix~$P=(p_{ij})_{i,j=1}^n$ with the following properties (in conditions given below, indices $i,j$ belong to $\{1,\ldots,n\}$):
\begin{equation}
\label{one}
\begin{array}{c}
\text{$0\le p_{ij}\le 1$ for all $i,j$};\\
\text{$\sum_{j=1}^n p_{ij}\le 1$ for all $i$ (the substochasticity);}\\
\text{the matrix $P$ is not nilpotent;}\\
\text{for any principal submatrix of the matrix~$P$}\\
\text{there exists a row such that the sum of its elements}\\ 
\text{in this submatrix is strictly less than one.}
\end{array}
\end{equation}
Note that with $P\equiv P_G$ the latter property is equivalent to the nonrecurrence of all states (except the absorbing one)~\cite{feller}; the matrix $P$ is nilpotent if and only if the graph $G$ is acyclic.

Recall that for matrices with nonnegative elements (\textit{nonnegative} matrices) the next theorem~\cite[Theorem~3, Chapter~XIII]{gantmaher} is valid. Namely, ``A non-negative matrix $A=(a_{ij})_{i,j=1}^n$ always has a non-negative characteristic value~$r$ such that moduli of all characteristic values of~$A$ do not exceed~$r$. To this {\it maximal} characteristic value~$r$ there corresponds a non-negative characteristic vector $Ay=ry$ ($y\ge 0$, $y\ne 0$).'' Note that both the matrix~$A$ and that $A^t$ (the symbol $t$ is the transposition sign) may have no {\it positive eigenvector} (a vector all whose components are strictly positive). Later we discuss existence conditions for such a vector.

Recall that the symbol $P(\beta)$ denotes the matrix $(p_{ij}^\beta)_{i,j=1}^n$ (here $0^\beta=0$ for any~$\beta$), while $G$ stands for a directed graph with $n$ vertices, whose arcs correspond to nonzero elements of the matrix~$P$.

\begin{mylemma}
\label{tone}
For any matrix $P$ in form~(\ref{one}) there exists unique $\beta\in\mathbb{R}$ such that the maximal characteristic value of the matrix $P(\beta)$ equals~1, while $0\le\beta < 1$. The inequality $\beta>0$ is equivalent to the existence in the graph $G$ of two different simple cycles that go through one and the same vertex.
\end{mylemma}

{\bf Proof of Lemma~\ref{tone}:} Denote by $s_i$ the sum $\sum_{j=1}^n p_{ij}$. Let $s=\min_i s_i$ and $S=\max_i s_i$. It is known that~\cite[Remark on p.~68]{gantmaher} the maximal characteristic value~$r$ of any nonnegative matrix satisfies the inequality~$s\le r\le S$. Denote by $r(\psi)$ (here $\psi\ge 0$)  the maximal eigenvalue of the matrix~~$P(\psi)$, let $s(\psi)=\min_i \sum_{j=1}^n p_{ij}^\psi$ and $S(\psi)=\max_i\sum_{j=1}^n p_{ij}^\psi$.

Let us prove the uniqueness of the choice of $\beta$ from the lemma condition and the validity of the inequality $0\le \beta<1$. Recall that the matrix~$P$ is called indecomposable if the oriented graph~$G$ is strongly connected. It is known that~\cite[p.~63]{gantmaher} indecomposable nonnegative matrices with unequal values of $s$ and $S$ satisfy the strict inequality $s<r<S$. In a general case, the decomposition of a graph into strongly connected components corresponds to the normal form of the matrix obtained from the initial one by renumbering its rows (and, correspondingly, columns). In the normal form (see~\cite[p.~75]{gantmaher}) the diagonal is occupied by square blocks corresponding to numbers of vertices that belong to one and the same strongly connected component; the matrix elements located above these blocks equal zero. Therefore, sequentially decomposing the determinant by the group of rows that correspond to strongly connected components, we obtain that the characteristic polynomial of the matrix~$P(\psi)$ equals the product of characteristic polynomials of each of diagonal blocks, $r(\psi)$ coincides with the maximal eigenvalue of blocks. However, according to formula~(\ref{one}), for square submatrices that correspond to each of these blocks, the value $s$ is strictly less than one. In addition, not all blocks are zero, otherwise the matrix~$P$ is nilpotent and $s(0)\ge 1$ for at least one of blocks. Consequently, $r(1)<1$ and $r(0)\ge 1$.

Evidently, $p_{ij}^\psi$ decreases as $\psi$ increases, if $ p_{ij}>0$. It is known that~\cite[Theorem~6, Chapter~XIII]{gantmaher} if some elements of a nonnegative indecomposable matrix decrease, then its maximal characteristic value strictly decreases. Therefore $r(\psi)$ is a decreasing function. We have proved the uniqueness of the choice of $\beta$ and the validity of the inequality $0\le \beta<1$.

Let us prove the last assertion of the lemma. In the normal form of the matrix~$P$ we consider the block containing the vertex that belongs to two different cycles. For this block we introduce analogs of values $s(\psi)$ and $S(\psi)$; we denote them by $s'(\psi)$ and $S'(\psi)$, correspondingly. The considered block, by definition, is an indecomposable matrix. Consequently, $s'(0)\ge 1$ and $S'(0)\ge 2$. Hence for the matrix $P(0)$ we get $r(0)>1$, which implies that in this case the desired value of $\beta$ (by condition of the lemma) is strictly positive.

It remains to prove that if no vertex in the graph~$G$ belongs to two cycles, then the desired value of $\beta$ equals zero. Really, the considered diagonal blocks either are trivial (i.e., consisting of one element) or correspond to nontrivial strongly connected components of the graph~$G$. A nontrivial component, by definition, contains a cycle going through all its vertices. In our case this cycle cannot be self-intersecting, because in this case there would exist a vertex belonging to two cycles. For the same reason, there are no arcs different from those of the considered (simple) cycle in the strongly connected component. But this means that for the corresponding block, $S'(0)=s'(0)=1$. Since the characteristic polynomial of the matrix $P(0)$ represents the product of characteristic polynomials of diagonal blocks, we obtain $r(0)=1$. The lemma is proved.

\begin{myso}
\label{oneSo}
Assume that under conditions of Lemma~\ref{tone}, $\beta>0$ and the normal form of the matrix~$P$ contains several blocks representing strongly connected components~$H$ of the graph~$G$ such that characteristic numbers of matrices $P_H(\beta)$ equal one. Then each of these graphs~$H$ contains a vertex that belongs to two (or more) different simple cycles.
\end{myso}

Evidently, Lemma~\ref{tone}, taking into account the nonrecurrence of states of the Markov chain, implies the existence of the exponent~$\beta$ in the interval~$(0,1)$, provided that conditions of Theorem~\ref{main}.2 are fulfilled.

Let us now consider the case when the matrix~$P(\beta)^t$ has a positive eigenvector corresponding to the unit eigenvalue. Redefining the standard necessary and sufficient conditions for the existence of a positive eigenvector (see~\cite[theorem~7, chapter~XIII]{gantmaher}), we obtain the following assertion.

\begin{myprop}
\label{propOne}
Let assumptions of Lemma~\ref{tone} be fulfilled and $\beta>0$. The matrix $P(\beta)^t$ has a positive eigenvector corresponding to the unit eigenvalue if and only if in the graph~$G'$ vertices without incoming arcs, and only they, correspond to strongly connected components~$H$, for which matrices $P_H(\beta)$ have the unit characteristic value.
\end{myprop}

\begin{myso}
\label{twoSo}
Assume that under conditions of Theorem~\ref{main}.2 the matrix $P(\beta)^t$ has a positive eigenvector corresponding to the unit eigenvalue. Then we can choose a vector $a=(a_1,\ldots,a_n)$ such that $a_k=0$ for all vertices with less than two incoming arcs, and the probability of reaching any vertex is greater than zero.
\end{myso}

{\bf Proof of Corollary~\ref{twoSo}:} Consider graphs~$H$ mentioned in Proposition~\ref{propOne}. According to Corollary~\ref{oneSo}, in each of them there exists a vertex which belongs to two cycles. Assume that $a_v>0$ for all such vertices~$v$, and $a_v=0$ otherwise. Then the probability to reach any vertex of graphs~$H$ is greater than zero, because all these vertices are located in one and the same strongly connected component. Proposition~\ref{propOne} implies that all the rest strongly connected components are also reachable with nonzero probabilities. But then we can get with nonzero probabilities to all vertices of the graph~$G$, which was to be proved.

\section{The power law in the case of the existence of a positive eigenvector.}

We need some more auxiliary assertions about power inequalities for the function~$p(t)$. Note that Lemma~2 is valid even without assumptions on the existence and positiveness of the eigenvector of the matrix $P(\beta)^T$. We use it for proving both the main result of this section (in the framework of the mentioned assumption), and its corollaries (in a more general case).

\begin{mylemma}
\label{ttwo}
{\rm\bf A.}~Let $\delta>0$. With some initial distribution~$a$ (not necessarily satisfying condition~(\ref{cond})) we obtain $p_{a}(t)=\Omega(t^{-\delta})$ (hereinafter the subscript indicates the initial distribution under consideration). Then with any initial distribution~$a'$ satisfying condition~(\ref{cond}) we have $p_{a'}(t)=\Omega(t^{-\delta})$.

{\rm\bf B.}~Let $\delta>0$. Assume that with some initial distribution~$a$, $a=(a_1,\ldots,a_n)$, satisfying condition~(\ref{cond}) it holds $p_{a}(t)=O(t^{-\delta})$. Then with any initial distribution~$a'$ we have $p_{a'}(t)=O(t^{-\delta})$.
\end{mylemma}

As a corollary, we obtain that if $p_{a}(t)=\Theta(t^{-\delta})$ with some initial distribution~$a$ satisfying~(\ref{cond}), then it is also valid for all initial distributions satisfying~(\ref{cond}).

{\bf Remark~4.} If the order of the asymptotics is not power, then the assertion analogous to Lemma~\ref{ttwo}, generally speaking, is not true. Namely, the order of the asymptotics of the function~$p(t)$, possibly, depends on the initial distribution. Thus, when calculating the Markov chain that corresponds to the (last) diagram~e in Fig.~1, we obtain the exponential order of the asymptotics of the function~$p(t)$ with the exponent $\nu=\ln\sqrt{2}$. Here we assume that $a_1>0, a_2>0$. But if $a=(1,0)$ in this chain, then, as one can easily prove, the asymptotics is exponential with $\nu=\ln 2$.

\medskip

In the proof of Lemma~\ref{ttwo} instead of the function $p(t)$ we consider the ``inverse'' to it function $Q(q)$, $q\in(0,1]$ (which equals the number of words whose probabilities are not less than~$q$). This assertion is equivalent to an analogous one for $Q(q)$ with $1/\delta$ in place of $\delta$. Really, the graph of the function $p(t)$ demonstrates that the inequality $p(t)<c\,t^{-\delta}$ ($p(t)>c\,t^{-\delta}$) with all $t\ge 1$ is equivalent to $Q(q)<(q/c)^{-1/\delta}=\text{\rm const}\,q^{-1/\delta}$ (or, respectively, $Q(q)>\text{\rm const}\, q^{-1/\delta}$) with all (sufficiently small) values of~$q$. 

We denote a Markov chain with an initial distribution~$a$ by $\text{MCh}_{a}$, we do probabilities of words $w$ in this Markov chain by $\mathop{\rm Pr}_{a}(w)$. By definition, all words in $\text{MCh}_{a}$ begin in the set $E(a)=\{E_i: a_i>0\}$, and we denote the corresponding set of vertices by $I(a)=\{i: a_i>0\}$. The idea of the proof consists in associating words in $\text{MCh}_{a}$ with those in $\text{MCh}_{a'}$, and then in estimating the function~$Q$.

{\bf Proof of Lemma~\ref{ttwo}.A:}
Evidently, for each $j$, $j\in I(a)$, there exists some path $(i',i_1,\ldots,j)$ such that $i'\in I(a')$; we denote this path by~$\pi(j)$. We associate each word~$w$ in $\text{MCh}_{a}$, beginning with $E_{j}$, with a word~$w'$ in $\text{MCh}_{a'}$ by adding the prefix $(E_{i'},E_{i_1},\ldots,E_{j})$. Evidently, $\mathop{\rm Pr}_{a'}(w')= \mathop{\rm Pr}_{a}(w) c(j)$, where $c(j)=\widetilde{\mathop{\rm Pr}}(\pi(j)) a'_{i'}/ a_{j}$. It is possible that several words in $\text{MCh}_{a}$ correspond to one and the same word in $\text{MCh}_{a'}$. However, in the associated list this word may appear no more than~$n$ times, because there exists no more than $n$ variants of prefixes that begin with $E_{i'}$.

Consider the sorted list of first $t$ words~$(w_1,w_2,\ldots,w_t)$ in $\text{MCh}_{a}$ and associate them with words~$(w'_1,\ldots,w'_t)$ in $\text{MCh}_{a'}$ (some of them, possibly, coincide). We get $p_{a'}(t)\ge \mathop{\rm Pr}_{a'}(w'_{nt})\ge p_{a}(n t)\min_{j\in I(a)} c(j)>\text{\rm const}\, t^{-1/\delta}$, and Lemma~\ref{ttwo}.A is proved.

Proof of Lemma~\ref{ttwo}.B is quite similar (it uses the inequality $p_{a'}(t)\le c\, p_{a}(\lceil t/n \rceil)$). 

Let us now prove the key lemma including an important particular case of Theorem~\ref{main}.2.

\begin{mylemma}
\label{tthree}
Assume that a graph~$G$ has a vertex that belongs to two different simple cycles, $\beta$ is chosen in accordance with Lemma~\ref{tone}, and the matrix $P(\beta)^t$ has a positive eigenvector~$e$ corresponding to the unit eigenvalue. Then $p(t)=\Theta(t^{-1/\beta})$.
\end{mylemma}

{\bf Proof} (cf. the proof in~\cite{VovaAndI}):

As was noted earlier (before the proof of Lemma~\ref{ttwo}), the assertion about the power asymptotics of the function~$p(t)$ is equivalent to an analogous assertion for the function~$Q$. Let us prove it now.

We understand an incomplete word as the initial part of a word (a path) $(i_1,\ldots,i_m)$ such that $a_{i_1}>0$; we define the ``probability'' of an incomplete word by the same formula~(\ref{PrWord}). For positive~$x$ we introduce functions $Q_k(x)$, $k=1,\ldots,n)$, which equal the number of incomplete words ending with the symbol $E_k$ whose ``probabilities'' are not less than $x$. Evidently, $Q_k(x)=0$ with $x>1$. We also need functions $\tilde{Q_k}(x)$: $\tilde{Q_k}(x)= Q_k(x)+1,$ $k=1,\ldots,n$.

Let us prove that $Q_k(x)=\Theta (x^{-\beta})$ as $x\to 0$. Evidently, such power estimate from above (from below) for the function $Q_k(x)$ is equivalent to an analogous estimate for $\tilde Q_k(x)$.

Put 
$$
\chi_0(x)=\left\{
\begin{matrix}1 &\text{ for } x \le 1,\\
0&\text{ for  } x>1.
\end{matrix}
\right.
$$
The definition implies the following important recurrent correlation:
\begin{eqnarray*}
& Q_k(x)=\sum_{m: p_{mk}>0} Q_m(x/p_{mk})+\chi_k(x),&\\ 
&\text{ where } \chi_k(x)=\left\{
\begin{matrix}\chi_0(x/a_k), &\text{$a_k>0$}, \\
0,&\text{otherwise}.
\end{matrix}
\right.&\nonumber
\end{eqnarray*}
In particular, the following inequality is valid:
\begin{equation}
\label{ge}
Q_k(x)\ge \sum_{m: p_{mk}>0} Q_m(x/p_{mk}),\quad k=1,\ldots,n.
\end{equation}

Let us now use Lemma~\ref{ttwo}, which gives some freedom of the choice of the initial distribution. Choosing $a_k$ as is described in Corollary~\ref{twoSo}, for all vertices~$k$ with one incoming arc $(m,k)$ we get $Q_k(x)=Q_m(x/p_{mk})$. But if the number of incoming arcs is less than two, then, evidently, $Q_k(x)\le \sum_{m: p_{mk}>0} Q_m(x/p_{mk})+(l-1)$, where $l$ is the number of terms in the sum. Therefore,
\begin{equation}
\label{le}
\tilde Q_k(x)\le \sum_{m: p_{mk}>0} \tilde Q_m(x/p_{mk}),\quad k=1,\ldots,n.
\end{equation}

Let the vector~$e$ mentioned in the condition of the lemma have components $(e_1,\ldots,e_n)$. One can easily make sure that functions $f_k(x)=e_k x^{-\beta}$, $k=1,\ldots,n$, satisfy the following set of functional equations:
\begin{equation}
\label{eq}
f_k(x)= \sum_{m: p_{mk}>0} f_m(x/p_{mk}),\quad k=1,\ldots,n.
\end{equation}

Now let $M$ be the minimum of positive elements of the matrix~$P$, and let $M'$ be the maximum of its non-unit elements. Fix $y$ such that $Q_k(y)>0$ for all $k$. Evidently that on the segment $[My,y]$ the function $Q_k(y)$ is monotone and positive (more exactly, on this segment it takes on a finite number of natural values). This means that one can find positive constants $c_1$ and $c_2$ independent of $k$ such that inequalities $Q_k(x)\ge c_1 f_k(x)$ and $\tilde Q_k(x)\le c_2 f_k(x)$, $k=1,\ldots,n$, are valid with $My\le x\le y$. But then formulas~(\ref{ge},\ref{le},\ref{eq}) imply that the same inequalities (with the same constants $c_1$ and $c_2$) are valid with $x\in [M'My,y]$ and, consequently, with all $x\le y$. The estimate $Q_k(x)=\Theta(x^{-\beta})$ for $x\le y$ is proved.

Since $Q(x)= \sum_{m: p_{m0}>0} Q_m(x/p_{m0})$, we obtain that $Q(x)=\Theta(x^{-\beta})$ for sufficiently small~$x$.

\begin{myso}
\label{threeSo}
Let assumptions of Theorem~\ref{main}.2 be fulfilled. Then $p(t)=\Omega(t^{-1/\beta})$, where $\beta$ is a real number such that the maximal modulo eigenvalue of the matrix $P_G(\beta)$ equals one.
\end{myso}

{\bf Proof of Corollary~\ref{threeSo}:} The idea of the proof consists in the application of Lemma~\ref{ttwo}.A. But first we need to find at least one initial distribution, for which our power estimate from below is valid.

Consider $\beta$ defined in the condition of Corollary~\ref{threeSo} (recall that in view of Lemma~\ref{tone} it exists and is positive and unique). In the normal form, the matrix $P_G(\beta)$ has blocks that represent strongly connected components~$H$ such that the maximal modulo eigenvalue of the matrix $P_H(\beta)$ equals one. Assume that conditions of Proposition~\ref{propOne} are violated. Then in some path in the graph~$G'$ one of such blocks does not correspond to the first vertex in the path. Without loss of generality, we can assume that no arc enters the initial vertex of the path under consideration. We delete this vertex from the graph~$G'$ and do the corresponding connectivity component from the graph~$G$. Consider the ``truncated'' Markov chain with the obtained graph. Evidently, as above, it satisfies conditions~(\ref{MC}) and assumptions of Theorem~\ref{main}.2; moreover, for the matrix of transition probabilities, the value of $\beta$ remains the same. 

Repeating this operation several times, we can make the matrix of the obtained graph $\tilde G$ satisfy conditions of Proposition~\ref{propOne}. Fixing the initial distribution~$a$ for the Markov chain with the graph~$\tilde G$, we fix some distribution~$a$ for the Markov chain with the graph $G$; however, in this case we never reach deleted vertices. By applying Lemma~\ref{tthree} (which is proved aove) and using Lemma~\ref{ttwo}.A, we obtain the assertion of Corollary~\ref{threeSo}.

\begin{myso}
\label{fourSo}
Let conditions of Theorem~\ref{main}.2 be fulfilled. Then $p(t)=o(t^{-1/\beta'})$ for any $\beta'>\beta$.
\end{myso}

{\bf Proof:} The idea of the proof consists in the application of Lemma~\ref{ttwo}.B. But first we perform the operation opposite to that in the proof of the previous lemma. Namely, we add to the graph $G$ additional strongly connected components so as to make the obtained Markov chain satisfy the condition of Lemma~\ref{tthree} with some exponent $\beta''$ lesser than~$\beta'$.

Let $k$ be the number of vertices in the graph $G'$ which have no incoming arcs, let $v$ be one of such vertices, and let $H_v$ be the corresponding to it strongly connected component of the graph~$G$. Let us add to $G$ some subgraphs~$\tilde H_v$ which have the form shown in the upper part of diagram~b), then an arc from the added subgraph will lead to one of vertices in~$H_v$. As a result, we will obtain a graph with $n+2k$ vertices. 

Consider a Markov chain with $n+2k$ non-absorbing states, whose matrix of transition probabilities $\tilde P$ is obtained from the matrix $P$ by adding $k$ pairs of rows that correspond to subgraphs~$\tilde H_v$. Each pair corresponds to a diagonal $2\times 2$ block in the form $P_2=\left(\begin{array}{cc}r&s\\t&0\end{array}\right)$, where $0<r,s,t<1$, $r+s=1$, numbers $r, s, t$ are the same for all blocks. Let us choose numbers $r, s, t$ so as to make the maximal eigenvalue of the matrix $P_2(\beta'')$ equal one (for some $\beta''$: $\beta<\beta''<\beta'$). To this end, it suffices to choose~$x$ such that $r^{\beta''}x+ s^{\beta''}=1$ (since $r^{\beta''}+ s^{\beta''}>1$, the desired value of $x$ is less than one), and then to set $t=x^{1/\beta''}$.

Let us now consider the Markov chain with the transition probability matrix (between non-absorbing states)~$\tilde P$. Evidently, the matrix $\tilde P(\beta'')$ satisfies conditions of Proposition~\ref{propOne}, whence by Lemma~\ref{tthree} and Lemma~\ref{ttwo}.B we get $p_a(t)=O(t^{-\beta''})$ for any initial distribution $a$ of this Markov chain. In particular, this is also valid for $I(a)\in V(G)$, and in this case we never reach vertices of added graphs $\tilde H(v)$. Thus, for the initial Markov chain we have $p(t)=O(t^{-\beta''})$, which was to be proved.

\section{Completion of the proof of Theorem~\ref{main}.2.}

It remains to establish necessary and sufficient conditions for the power asymptotics. Sufficient but not necessary conditions are given by assumptions of Lemma~\ref{tthree}. In order to complete the proof of Theorem~\ref{main}.2 with the help of Lemma~\ref{tthree}, we need two more auxiliary assertions.

Let us first consider the case of a ``parallel'' connection of graphs $G_1$ and $G_2$ of Markov chains (we denote the Markov chains themselves by $\text{MCh}_{G_1}$ and $\text{MCh}_{G_2}$); we identify the absorbing states of these graphs.

\medskip

\begin{figure}
\begin{center}
\includegraphics[width=0.45\textwidth]{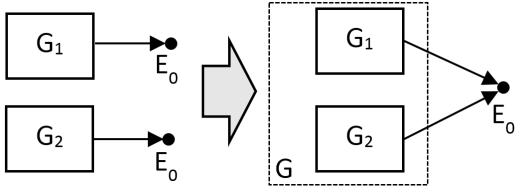}\quad
\caption{The construction of $\text{MCh}_G$ by the ``parallel'' connection of graphs of $\text{MCh}_{G_1}$ and $\text{MCh}_{G_2}$. Arcs that earlier led from $G_1$ and $G_2$ to their ``own'' absorbing states, now lead to the common absorbing state $E_0$.}
\end{center}
\label{Pic3}
\end{figure}

\begin{mylemma}
\label{tfour}
Assume that Markov chains with graphs~$G_1$ and $G_2$ with some initial distributions (satisfying condition~(\ref{cond})) for $p_1(t)$ and $p_2(t)$ (probabilities of the $t$th word in the corresponding sorted list) satisfy correlations $p_1(t)=O(t^{-\delta_1})$ and $p_2(t)=O(t^{-\delta_2})$, where $\delta_1, \delta_2 >0$. Assume that for the Markov chain with the function $p(t)$, any word represents either a word from the first Markov chain or that of the second one; its graph $G$ represents a non-connected union of graphs $G_1$ and $G_2$, while the corresponding transition probabilities remain the same (see Fig.~3). Then with any initial distribution the following correlation is valid:
\begin{equation}
\label{eneq}
p(t)=O(t^{-\delta}),\ \text{where $\delta=\min\{\delta_1,\delta_2\}$.}
\end{equation}
\end{mylemma}

{\bf Proof:} By Lemma~\ref{ttwo}.B it suffices to prove inequality~(\ref{eneq}) with some concrete initial distribution~$a$ satisfying condition~(\ref{cond}). Let us choose it as $(a'+a'')/2$, where $a'$ and $a''$ are initial probability distributions in the first and second Markov chains, correspondingly. Then probabilities of all words in the aggregated Markov chain are 2 times less than probabilities of the same words in calculations of $p_1(t)$ and $p_2(t)$. The list of first $t$ words of our Markov chain sorted in the non-increasing order of their probabilities consists of the initial part of the analogous list of the first MCh alternated with the initial part of the second MCh; consequently, this list contains a word of either first or second MCh with the index $\lceil t/2 \rceil$. We have
\begin{equation}
\label{t2}
p(t)\le \max\{p_1(\lceil t/2 \rceil), p_2(\lceil t/2 \rceil )\}
\end{equation}
(we could have again divide the right-hand side by 2, but even the weakened variant of the inequality suits us).

By condition there exist positive constants $c_1$ and $c_2$ such that
\begin{equation}
\label{c1c2}
p_1(t)<c_1\, t^{-\delta_1},\quad p_2(t)<c_2\,t^{-\delta_2}\ \text{for all~$t$.}
\end{equation}
Let us choose a constant $c$ such that
$c\,t^{-\delta}>\max\{2^{\delta_1} c_1 t^{-\delta_1},2^{\delta_2} c_2 t^{-\delta_2}\}$ for all natural~$t$. Using~(\ref{t2}) and (\ref{c1c2}), we obtain $p(t)< c\,t^{-\delta}$.

\medskip
{\bf Remark~5.} Evidently, Lemma~\ref{tfour} can be extended by induction to the case of the ``parallel'' connection of $\text{MCh}_{G_1}, \text{MCh}_{G_2},\ldots,\text{MCh}_{G_m}$.
\medskip

Let us now consider the case when graphs of Markov chains are connected ``sequentially''. Consider the graph $G$ obtained from the union of graphs~$G_1$ and~$G_2$ of Markov chains by redirecting at least some arcs that earlier led from~$G_1$ to the absorbing state, and now do to the graph~$G_2$. Denote the set of these arcs by~$E_{12}$. Assume that one can reach any vertex of the graph~$G_2$ along the path that goes through the proper arc from~$E_{12}$, and transition probabilities in $\text{MCh}_G$ are equal to the corresponding probabilities in $\text{MCh}_{G_1}$ and $\text{MCh}_{G_2}$ (see Fig.~4).

\medskip

\begin{figure}
\begin{center}
\includegraphics[width=0.45\textwidth]{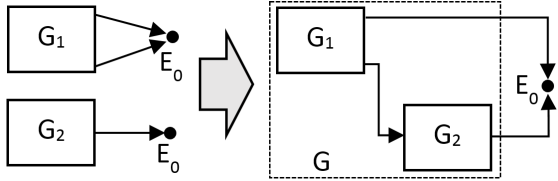}\quad
\caption{The construction of $\text{MCh}_G$ by a ``sequential'' connection of graphs of $\text{MCh}_{G_1}$ and $\text{MCh}_{G_2}$. Arcs that earlier led from $G_1$ to their ``own'' absorbing states form two groups; arcs of the first group lead to the common absorbing state $E_0$, those of the second one do to the graph $G_2$. All arcs that earlier led from $G_2$ to their ``own'' absorbing states now lead to the common absorbing state $E_0$.}
\end{center}
\label{Pic4}
\end{figure}

\begin{mylemma}
\label{tfive}
Assume that Markov chains with graphs~$G_1$ and $G_2$ with some initial distributions (satisfying condition~(\ref{cond})) for $p_1(t)$ and $p_2(t)$ (probabilities of the $t$th word in the corresponding sorted list) fulfill correlations $p_1(t)=O(t^{-\delta_1})$ and $p_2(t)=O(t^{-\delta_2})$, where $\delta_1, \delta_2 >0$. Let the Markov chain with the function $p(t)$ correspond to the graph~$G$ representing the union of graphs $G_1$ and $G_2$ with additional arcs going from the graph~$G_1$ to that~$G_2$ so that any vertex of the graph~$G_2$ is attainable through the path consisting of these arcs. Then formula~(\ref{eneq}) is valid with $\delta_1\ne \delta_2$. Correlation~(\ref{eneq}) is false if the initial distribution satisfies condition~(\ref{cond}), while $\delta_1=\delta_2$ and $p_1(t)=\Omega(t^{-\delta_1})$, $p_2(t)=\Omega(t^{-\delta_2})$.
\end{mylemma}

{\bf Proof of Lemma~\ref{tfive}:}
As the initial distribution in $\text{MCh}_G$ we consider a distribution~$a$ concentrated at vertices of the graph~$G_1$ and satisfying condition~(\ref{cond}) for it. Evidently, for $\text{MCh}_G$ condition~(\ref{cond}) is also valid; further considerations are related to the corresponding function~$p(t)$. 

Note that the assertion of Lemma~\ref{ttwo} remains valid, even if the probability~$a_0$ that the initial state is absorbing differs from zero. Moreover, in this case, in order to make the sum of probabilities of all words equal one, it is convenient to add to the sorted list of all possible words one more word, the empty one, whose probability equals~$a_0$ (this, naturally, does not affect the asymptotic properties of considered functions). 

Assume that the constant $c_1$ in inequality~(\ref{c1c2}) is defined for the initial distribution $a'$ in $\text{MCh}_{G_1}$ coinciding with the distribution~$a$. We assume that the initial distribution~$a''$ in $\text{MCh}_{G_2}$ is concentrated at end vertices~$E_{12}$ and at the absorbing state. Moreover, values $a''_i$ equal probabilities of reaching the corresponding states in $\text{MCh}_{G}$ with the initial distribution~$a$. Taking into account the remark in the previous paragraph, we assume that the constant~$c_2$ in inequality~(\ref{c1c2}) is defined just for the initial distribution~$a''$. In addition, if earlier $p_i(t)=\Omega(t^{-\delta_i})$, $i=1,2$, then we denote by $c'_1,c'_2>0$ constants such that $p_1(t)>c'_1 t^{-\delta_1}$ and $p_2(t)>c'_2 t^{-\delta_2}$. 

Instead of power estimates for the function $p(t)$, let us prove power estimates for the function $Q(q)$. Let us first consider the case $\delta_1\ne\delta_2$.

First of all, note that any word $w$ in the initial Markov chain is representable in the form $(w_1,w_2)$, where $w_i$, $i=1,2$, are words of the Markov chain with the graph $G_i$. Here, as one can easily see, $\mathop{\rm Pr}_G(w)=\mathop{\rm Pr}_{G_1}(w_1)\mathop{\rm Pr}_{G_2}(w_2)$ (the subscript at the symbol Pr indicates the graph of the Markov chain, where we consider the word).

Evidently, $\mathop{\rm Pr}_{G_i}(w_i)=p_i(t_i)$, where $t_i$ is the number of the word $w_i$ in the corresponding list. Assuming that $\delta_1>\delta_2$, we get (below $t_1,t_2$ run over all possible natural values):
\begin{eqnarray*}
& Q(q)=\left|\{(t_1,t_2): p_1(t_1)p_2(t_2)\ge q\}\right|\le & \\
& \le \left|\{(t_1,t_2): c_1 t_1^{-\delta_1} c_2 t_2^{-\delta_2}\ge q\}\right|=&\\
&=\left|\{(t_1,t_2): t_1^{\delta_1} t_2^{\delta_2}\le (c_1 c_2)/q\}\right|\le &\\
&\le \sum_{t_1=1}^{\infty}(q/(c_1 c_2))^{-1/\delta_2} t_1^{-\delta_1/\delta_2}=\text{const}\ q^{-1/\delta_2}.&\nonumber
\end{eqnarray*}

In the case $\delta_1=\delta_2=\delta$ analogous considerations lead to the inequality
$$
Q(q)\ge \left|\{(t_1,t_2): t_1 t_2\le ((c'_1 c'_2)/q)^{1/\delta}\}\right|.
$$

According to the Dirichlet formula for the divisor function~\cite[chapter~XII]{titchmarsh}, the number of points with natural coordinates, whose product does not exceed $N$, equals $N\ln N+(2\gamma-1)N+O(\sqrt{N})$, where $\gamma$ is the Euler constant. Therefore, the inequality $Q(q)\le \text{const}\ q^{-1/\delta}$ can be fulfilled with small~$q$ with no positive constant, which was to be proved.

\medskip

{\bf Completion of the proof of Theorem~\ref{main}.2:} 
We prove that $p(t)=\Theta(t^{-1/\beta})$ under conditions of Theorem~\ref{main}.2 by induction with respect to the length of the maximal path in the graph~$G'$. If the graph $G'$ consists of unconnected vertices, then the assertion of the Theorem follows from Remark~5 and Lemma~\ref{tthree}. Otherwise we represent the graph $G$ as a ``sequential'' connection of the graph~$G_1$ consisting of strongly connected components corresponding to initial vertices of the graph~$G'$ (vertices without incoming arcs), and the graph~$G_2$ consisting of the rest part of the graph~$G$. Applying Lemma~\ref{tfive} (and the induction hypothesis for the graph~$G_2$), we obtain $p(t)=O(t^{-1/\beta})$. Consequently (see Corollary~\ref{threeSo}), $p(t)=\Theta(t^{-1/\beta})$.

Let us prove the necessity of conditions for the power asymptotics in Theorem~\ref{main}.2. Assume the contrary. Consider a path in the graph $G'$ with exactly two vertices corresponding to graphs $H_1$ and $H_2$ for which $P_{H_1}(\beta)$ and $P_{H_2}(\beta)$ have unit characteristic values. We can choose $H_1$ such that any path in the graph~$G'$ beginning at~$H_1$ contains no more than one such vertex of~$H_2$. Really, otherwise there exists a path $G'$ beginning at $H_2$ that contains a vertex of $H_3$, where $P_{H_3}(\beta)$ has the unit characteristic value. Then we can choose for $H_1$ the former graph $H_2$ (and do $H_3$ for $H_2$), and so on till the desired condition is fulfilled.

Consider an initial distribution~$a$ (not necessarily satisfying conditions~(\ref{cond})) concentrated at vertices of the graph~$H_1$. Let~$\widetilde{G}$ be the part of the graph $G$ reachable from these vertices. According to Lemma~\ref{ttwo}, the necessity of conditions of the power order for $\text{MCh}_{\widetilde{G}}$ automatically implies its necessity for $\text{MCh}_{G}$. 

The graph $\widetilde{G}$ is representable as a ``sequential'' connection of the graph~$G_1\equiv H_1$ and the graph~$G_2$ consisting of the rest part of the graph~$\widetilde{G}$. As was proved above, for the graph~$G_2$ it holds $p_2(t)=\Theta(t^{-1/\beta})$. Analogous inequalities $p_1(t)=\Theta(t^{-1/\beta})$ for the graph $G_1\equiv H_1$ are proved in Lemma~\ref{tthree}.  Applying the final part of Lemma~\ref{tfive}, we conclude that conditions of the power order with the exponent $-1/\beta$ cannot be fulfilled for $\text{MCh}_{\widetilde{G}}$ and, consequently, for $\text{MCh}_{G}$.

\section{Proof of Theorem~\ref{main}.3.}

In this case nontrivial strongly connected components of the graph~$G$ represent the considered cycles, and the graph $G'$ is obtained by contracting these cycles. Denote by $c'$ the cycle $\mathop{\rm argmax}_{c\in C} \widetilde{\mathop{\rm Pr}}(c)$, $\alpha=\widetilde{\mathop{\rm Pr}}(c')$ (note that $\alpha<1$). Let $v$ be one of vertices of this cycle. Let $v'$ be the vertex of the graph~$G'$ corresponding to the cycle~$c'$.

By condition~(\ref{cond}) there exists a word~$w$ containing the state~$E_v$. We set $w^{(0)}=w$, and $w^{(t)}$ is the word obtained from $w^{(t-1)}$ by inserting in it the sequence of states that correspond to the tracing of the cycle~$c'$. Evidently, $\mathop{\rm Pr}(w^{(t)}) =c_1 \alpha^{t}$, where $c_1=\mathop{\rm Pr}(w)$. By definition, $p(t)\ge \mathop{\rm Pr}(w^{(t-1)})> c_1 \alpha^{t}$. The lower exponential bound is proved.

Let us now prove that $p(t)$ decreases faster than any power function. Let $W$ be the {\it set of all words with nonrepeating states}. Evidently, each word~$w$ can be obtained from some word~$w'$ in $W$ by insertion of cycles. Some of these cycles are, possibly, repeating, however, they have to be subsequent in the considered path (since the graph~$G'$ is acyclic, it is impossible that the path of the graph~$G$ first goes through some cycle~$c$, then it does through a part that has no common vertices with the cycle, and then there appear vertices of the same cycle~$c$). The order of nonrepeating cycles is defined by the word~$w'$. Note that the result of the insertion is independent of the state (the letter) after which a fixed cycle~$c$ is inserted in the word (for example, for the (last) diagram~e in Fig.\,\ref{lernerPic1} the insertion of the cycle $1 \leftrightarrows 2$ into the word $(E_1,E_2,E_0)$ after the ``letter''~$E_2$ or after the ``letter'' $E_1$ gives one and the same word $(E_1,E_2,E_1,E_2,E_0)$).

Note that any word, whose length exceeds~$n\tau$, contains at least~$tau-1$ cycles. Consequently, Proposition~\ref{propTwo} is valid.

\begin{myprop}
\label{propTwo}
If $L(w)>n\tau$, then $\mathop{\rm Pr}(w)< \alpha^{\tau-1}$.
\end{myprop}

If $L(w)>n(\tau+1)$, then by Proposition~\ref{propTwo} we have $\mathop{\rm Pr}(w)<\alpha^{\tau}$. We are interested in the upper bound for the number of words~$w$ such that $\mathop{\rm Pr}(w)\ge\alpha^{\tau}$. In order to obtain this bound, suffice it to calculate the total number of words whose length does not exceed~$n(\tau+1)$.

Let us prove that under assumptions of the theorem the number of words, whose length does not exceed $x$, is bounded from above by the value $|W|(x+n)^n/n$, where $|W|$ is the cardinality of the set $W$. Really, any word-path of length~$i$ contains no more than~$i$ cycles. Evidently, the graph~$G$ has no more than~$n$ different cycles. Since the number of combinations with repetitions from~$n$ by~$i$ equals~$\binom{n+i-1}{i}$, the total number of words of length~$i$ is bounded from above by the value $|W|\binom{n+i-1}{i}$. Summing with respect to~$i$ from $0$ to~$x$, we obtain $|W|(1+x) \binom{n+x}{n-1}/n$, which gives the desired value.

The obtained estimate implies that the number of words, whose length does not exceed $n(\tau+1)$, is bounded from above by the value $f(\tau)=|W|n^{n-1} (\tau+2)^n$. Comparing this assertion with Proposition~\ref{propTwo}, we conclude that with $t>f(\tau)$ the inequality $p(t)< \alpha^{\tau}$ is fulfilled. Therefore, $p(t)=O(\alpha^{\sqrt[n]{t}/n})$, which proves the correlation $p(t)=o(t^{-\lambda})$.

Let us now prove necessary and sufficient conditions for the exponential decrease. It suffices to prove that if a graph contains a path going through vertices of two cycles, then $p(t)=\Omega(\delta^{\sqrt{2t}})$ for some $\delta $. The idea of the proof is analogous to that used for establishing the exponential lower estimate at the beginning of this section.

Thus, assume that the graph~$G$ contains a path going through vertices of two cycles, namely, first it does through vertices of a cycle~$c''$ and then those of~$c'''$. According to~(\ref{cond}), there exists a word~$w'$ containing both these vertices-states in the same order. Denote by $w^{(\tau)}$ the word obtained from~$w'$ by inserting states corresponding to $\tau$ cycles, each of which is either the cycle~$c''$ or that~$c'''$. Note that there is~$\tau+1$ ways to obtain the word $w^{(\tau)}$; each way consists in a combination with repetitions from 2 by~$\tau$. For any $w^{(\tau)}$ we have $\mathop{\rm Pr} (w^{(\tau)})\ge \mathop{\rm Pr}(w')\delta^\tau$, where $\delta=\min\{\widetilde{\mathop{\rm Pr}}(c''), \widetilde{\mathop{\rm Pr}}(c''')\}$. Thus, with $t=(\tau+1)(\tau+2)/2\equiv 1+2+\ldots+\tau$ the value of the function $p(t)$ is bounded from below by the minimal probability of one of considered words $w^{(i)}, i=0,\ldots,t$, i.e., $p(t)=\Omega(\delta^\tau)=\Omega(\delta^{\sqrt{2t}})$.

Let us now prove the last assertion of the theorem about the constant~$\nu$. Consider the set of words with non-repeating states~$W$. Now each word~$w$ can be obtained from some word~$w'$, $w'\in W$, by inserting one and the same cycle (possibly, repeated several times). Here a concrete cycle $c\in C$ can be inserted only in $k(c)$ words. Denote the set of such words by~$K(c)$. Evidently, $\mathop{\rm Pr}(w)= \mathop{\rm Pr}(w')\widetilde{\mathop{\rm Pr}}(c)^m$, where $m$ is the number of cycles~$c$ inserted in the word $w'\in K(c)$.

Let $p'<\min_{w'\in W} \mathop{\rm Pr}(w')$. Let us find the number of words $Q(p')$ whose probabilities exceed~$p'$. Evidently, such are all words in~$W$. We can obtain the rest words by inserting some cycle $c$ in one of words from the set~$K(c)$. The number of inserted cycles varies from zero to $\lfloor(\ln p'-\ln \mathop{\rm Pr}(w'))/ \ln \widetilde{\mathop{\rm Pr}}(c) \rfloor$. Therefore, the difference $Q(p')- \ln p' \sum_{c\in C}k(c)/\ln \widetilde{\mathop{\rm Pr}}(c)$ is a bounded value. The proved boundedness of the difference $Q(\exp\{-\nu x\})-x$ with all $x>0$ is equivalent to the boundedness of the difference $t-\ln\{ 1/p(t)\}/ \gamma'$, which was to be proved.

\section*{Acknowledgment}

We are grateful to Yu.A.~Al'pin for remarks which were useful, in particular, for improving the proof in Section~3.

This work was supported by the Russian Fo\-undation for Basic Research, grant~N~12-06-00404-a.

\end{document}